\documentclass[12pt]{article}
\usepackage{jheppubmod,amsmath,amssymb}
\pdfoutput=1
\usepackage{color}
\usepackage{amsmath,amssymb}
\usepackage{comment}
\usepackage{braket}
\usepackage{mathtools}
\mathtoolsset{showonlyrefs}
\usepackage{psfrag}
\usepackage{array}
\usepackage{amssymb}
\usepackage{amsmath}
\usepackage{amsthm}
\usepackage{graphicx}
\usepackage{epstopdf}
	
\usepackage{color}
\usepackage{epsfig}
\usepackage[punctsep]{collref}
\newcommand{\op}{{\cal O}} %GFF operator on the right 

\def\al{A}
\def\asmall{{\cal A}_{\text{small}}}

\def\pb[#1,#2]{\{#1, #2\}}
\def\deb[#1,#2]{[#1,#2]_{\text{D.B.}}}

\def\tr{{\rm Tr}}

\def\tbb{{\cal B}}

\def\Or[#1]{{\text{O}}\left({#1}\right)}
\def\dotl[#1,#2]{\left\langle #1,\, #2 \right\rangle}
\def\dotlb[#1,#2]{\left\langle #1,\, #2 \right\rangle}
\def\dotlm[#1,#2]{\left[ #1,\, #2 \right]}
\def\dotp[#1,#2]{(\vect{#1} \cdot\vect{#2})}
\def\aff[#1,#2]{\hat{#1}(#2)}

\def\n4sym{{\cal N}=4 SYM}
\def\>{\rangle}
\def\<{\langle}
\def\weight[#1,#2,#3]{\{(#1),#2,#3\}}
\def\ads[#1]{$\text{AdS}_{#1}$}

\def\tarelr[#1]{\widetilde{a}^{\text{rel}}_{R#1}}
\def\Oright[#1]{{\cal O}_{R#1}}
\def\Oleft[#1]{{\cal O}_{L#1}}
\def\aleft[#1]{a_{L#1}}
\def\arelr[#1]{a^{\text{rel}}_{R#1}}

\def\tband{T_{\text{b}}}

\def\tband{\text{time band}}
\def\band{\text{band}}
\def\heft{{\cal H_{\rm EFT}}}

\newcommand{\be}{\begin{equation}}
\newcommand{\ee}{\end{equation}}
\newcommand{\ba}{\begin{align}}
\newcommand{\ea}{\end{align}}
\newcommand{\bs}{\begin{split}}
\def\sess\end{split}
\def\D{\cal D}
\def\tD{\overline{\cal D}}
\newcommand{\vect}[1]{{#1}}

\def\dmax{{\cal D}_{m}}
\title{A toy model of black hole complementarity}
\author[a]{Souvik Banerjee,}
\emailAdd{souvik.banerjee@rug.nl}
\affiliation[a]{Van Swinderen Institute for Particle Physics and Gravity, University of Groningen, Nijenborgh 4, 9747 AG, The Netherlands }
\author[a]{Jan-Willem Bryan,}
\emailAdd{j.w.a.brijan@rug.nl}
\author[b,a]{Kyriakos Papadodimas}
\emailAdd{kyriakos.papadodimas@cern.ch}
\affiliation[b]{Theoretical Physics Department, CERN, CH-1211 Geneva 23,
Switzerland}
\author[c]{and Suvrat Raju}
\emailAdd{suvrat@icts.res.in}
\affiliation[c]{International Centre for Theoretical Sciences, Tata Institute of Fundamental Research, Shivakote, Bengaluru 560089, India}

\abstract{We consider the algebra of simple operators defined in a time band in a CFT with a holographic dual. When the band is smaller than the light crossing time of AdS, an entire causal diamond in the center of AdS is separated from the band by a horizon.  We show that this algebra obeys a version of the Reeh-Schlieder theorem: the action of the algebra on the CFT vacuum can approximate any low energy state in the CFT arbitrarily well, but no operator within the algebra can exactly annihilate the vacuum. We show how to relate local excitations in the complement of the central diamond to simple operators in the band.  Local excitations within the diamond are invisible to the algebra of simple operators in the band  by causality, but can be related to complicated operators called ``precursors''. We use the Reeh-Schlieder theorem to write down a  simple and explicit formula for these precursors on the boundary. We comment on the implications of our results for black hole complementarity and the emergence of bulk locality from the boundary.}

\begin{document}

\maketitle
\section{Introduction}
\begin{figure}[!h]
\label{compsketch}
\begin{center}
\includegraphics[width=0.3\textwidth]{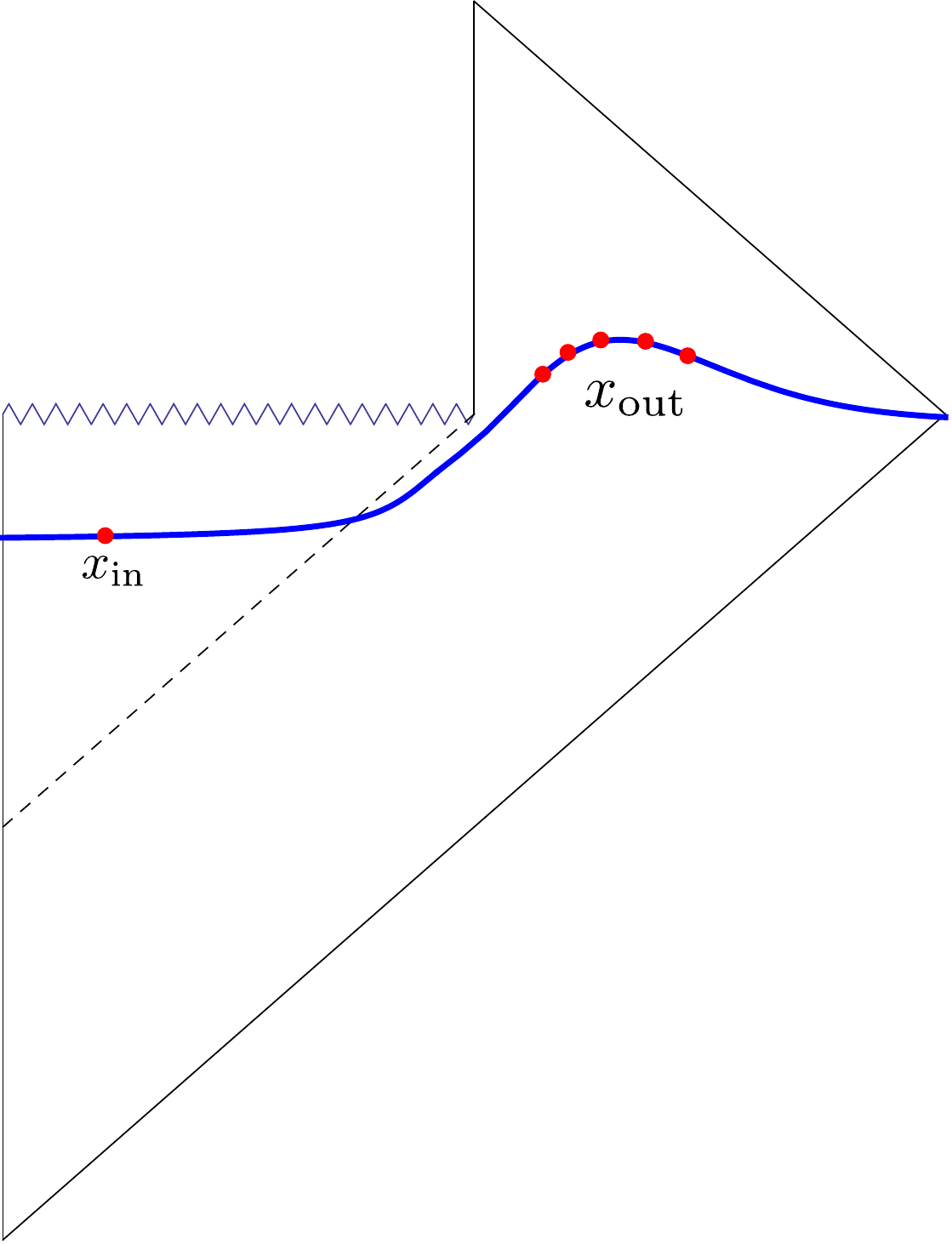}
\caption{\em Degrees of freedom at $x_{\rm in}$ are identified with complicated combinations of those at $x_{\rm out}$.}
\end{center}
\end{figure}

Recent discussions of the information paradox \cite{Mathur:2009hf,Almheiri:2012rt,Almheiri:2013hfa,Marolf:2013dba} have revived interest in the idea of black hole complementarity \cite{'tHooft:1984re,Susskind:1993if}. In a colloquial sense, this is the idea that degrees of freedom inside the black hole are ``scrambled'' combinations of degrees of freedom outside. A precise version of this idea was developed in \cite{Papadodimas:2013jku,Papadodimas:2013wnh,Papadodimas:2015jra,  Papadodimas:2015xma}. In this construction, a local operator inside the black hole can be represented as a sufficiently complicated combination of $\Or[S_{BH}]$ operators outside the black hole, where $S_{BH}$ is the black hole entropy. This is indicated by 
\be
\label{losslocality}
\phi(x_{in}) \cong  P(\phi(x_{2}), \phi(x_3), \ldots),
\ee
where $\phi(x_{in})$ is a local field located inside the black hole and $P$ is a suitably complicated polynomial comprised entirely of fields localized at points $x_{out} \equiv \{x_{2}, x_{3}, \ldots \}$ outside the black hole that are spacelike to $x_{in}$. (See figure \ref{compsketch}.)

It is apparent that such a relation implies a radical loss of causality and locality in correlators with $\Or[S_{BH}]$ insertions. However, the analysis of \cite{Papadodimas:2013jku,Papadodimas:2013wnh,Papadodimas:2015jra, Papadodimas:2015xma} left two questions unanswered. First, what is the precise form of $P$ that we must pick in order to observe this loss of locality, and in what sense is the field inside equal to the polynomial? Second, while positing such a loss of locality resolves various aspects of the information paradox, what is the independent evidence that this is indeed a physical feature of quantum gravity?

The objective of this paper is to investigate these questions in a simpler setting. We describe, in the context of the AdS/CFT correspondence  \cite{Maldacena:1997re,Witten:1998qj,Gubser:1998bc}, how such a loss of locality is not only physical but can even  be seen in empty AdS in the absence of black holes. In particular, given a local field operator $\phi(x_1)$, we will explicitly find ``complicated'' polynomials, ${\cal P}$ made up of field operators with support on points that are all spacelike with respect to $x_1$, but with the property that $\phi(x_1) \doteq \cal P$. In this equation the symbol $\doteq$ means that we can approximate $\phi(x_1)$ as accurately as we wish, but true equality can only be obtained by taking the limit of an infinite sequence of polynomials. This makes \eqref{losslocality} precise and thus demonstrates non-locality in a calculable setting.

Our setup in this paper is as follows. We consider a large $N$ CFT with a bulk AdS dual. In this CFT, we consider the set of boundary operators defined in a short $\tband$. Now, if this time band is shorter than the light-crossing time of AdS, it naturally divides the bulk spacetime into two regions: a causal diamond near the center of AdS that is causally disconnected from the time band, and its complement. (See figure \ref{diamond}.)

We argue that the CFT dual of this division is that operators in the CFT  can be naturally divided into two classes: simple operators ---which consist of single trace operators and polynomials of an $\Or[1]$ number of single trace operators ---  and complex operators, where the number of single-trace components starts scaling as a function of $N$.    While the region of AdS near the boundary is reconstructed by simple operators, we argue that  the region near the center of AdS 
is reconstructed by complicated ones. The two sets approximately commute in the large $N$ limit and on a given class of states. However, their commutator is not zero as an operator equation, and in fact, complicated operators can in principle be reconstructed by many simple operators. 
  
To demonstrate this feature, we identify the specific complicated operators that probe the diamond at the center of AdS. It turns out that the key ingredient we need for these operators are the polynomials
\be
\label{complicatedpoly}
{\cal P}_{\alpha, p_c} = \sum_{p=0}^{p_c} (-1)^p {(\alpha H)^p \over p!},
\ee
specified by two adjustable cutoffs, $\alpha, p_c$ that we can choose freely, provided we take them to be large enough. For example, as we describe below, one sufficiently large choice is  $\alpha = \ln (N)$ and $p_c =  N \ln(N)$. Here $H$ is the CFT Hamiltonian, shifted appropriately so that the ground state has energy $0$. Note that 
\be
\lim_{\alpha \rightarrow \infty} \lim_{p_c \rightarrow \infty} {\cal P}_{\alpha, p_c} = P_0,
\ee
 where $P_0 = |0\rangle \langle 0|$ is the projector onto the CFT vacuum. 
 We show that combining these complicated polynomials with other simple polynomials of single-trace operators allows us to probe regions which, in the bulk, are causally disconnected from the $\tband$. 

These observations demonstrate several important physical points: in order to understand locality in AdS/CFT, we have to  distinguish between simple and complicated experiments. Since effective field theory only requires locality to hold for simple experiments, we can indeed have significant violations of locality once we consider  complicated operators in the theory. 

A second important point is the relation between radial depth, complexity and time-dependence. The fact that we have locality in the emergent radial direction
in AdS/CFT is related to the fact that at any given moment in time, the information of the quantum state of the CFT is partly contained in simple and partly
in complicated operators. It is hard for a boundary observer to extract the information in the complicated operators, which is geometrically related to
the fact that she does not have direct access to points deep in AdS. Under time evolution the dynamics of the CFT shuffles the information between simple and complex operators. For states near the vacuum of AdS, the shuffling time is of the order of the AdS light crossing time. 

Hence complicated
operators inside a short $\tband$ can become simple operators at later points outside the $\band$. On the other hand, if the state corresponds
to a black hole in the bulk, then the information gets trapped in complicated operators for a very long time, until it manages to escape via Hawking evaporation. 

It is remarkable that the study of CFT operators in the short-time band can also capture some of the essential physics of black hole complementarity. We believe that this model deserves further attention. 

Before we close this section, we would like to mention that the same setup of a spherical hole in AdS was first considered in 
\cite{Balasubramanian:2013lsa}. (See  \cite{Myers:2014jia, Headrick:2014eia} for related work.) There it was proposed
that the decomposition of the bulk into $\D$ and $\tD$ could be understood in the CFT by covering the $\tband$ ${\cal B}$ with a set of overlapping causal
diamonds, whose size was determined by that of ${\cal B}$,  and considering the information that could be recovered by localized measurements in these diamonds. The proposal in this paper differs in that it concentrates on the decomposition of the boundary algebra into simple and complicated operators.

Second, this work has overlap with questions discussed in \cite{Marolf:2008mg, Marolf:2008mf,  Mintun:2015qda, Almheiri:2014lwa, Giddings:2015lla}. The ``code subspace'' introduced in  \cite{Almheiri:2014lwa} is similar to the subspace created by acting with the small algebra on the ground state of the CFT. As noted there, this construction was, in turn, related to a similar subspace---termed  ${\cal H}_{\Psi}$, and obtained by acting with the small algebra on an equilibrium black hole state --- that played a role in the reconstruction of the black hole interior \cite{Papadodimas:2012aq, Papadodimas:2013jku,Papadodimas:2013wnh,Papadodimas:2015jra,  Papadodimas:2015xma}. However, we do not consider the question of the spatial
localization of the information in the CFT and the possible connection to quantum error correction that was a central part of the discussion in \cite{Almheiri:2014lwa}.

There is also significant discussion in the literature on reconstructing the bulk from a subregion on the boundary that does have a causal complement. It was proposed in \cite{Headrick:2014cta} that this region is dual to bulk region called the ``entanglement wedge'', and we refer the reader to  \cite{Jafferis:2015del,Dong:2016eik,Freivogel:2016zsb} for some recent work on this proposal.  The Reeh-Schlieder theorem has also been previously considered in the context of AdS/CFT in the paper \cite{Morrison:2014jha} that considered the algebra of operators on the boundary of bulk wedges. 

Our emphasis in this paper is somewhat different from the papers above   because we are considering an entire band on the boundary. The ``entanglement wedge'' dual to this contains an entire Cauchy slice for the bulk. The reason we nevertheless are able to define a consistent subalgebra on the boundary is because of our division of boundary operators into simple and complicated operators.

This paper is organized as follows. In section \ref{secsetup} we describe our setup in more detail. In section \ref{secopalgebra} we describe the division of this algebra into simple and complicated operators in a more precise manner. We also prove a version of the Reeh-Schlieder theorem for operators confined to a finite band in time. In section \ref{secinterioroperators}, we describe how operators near the center of AdS that are causally disconnected from the time band, can nevertheless be constructed in terms of operators in the time band using suitably complicated operators. Some additional implications are discussed in section \ref{secconclusion}.

\section{The setup \label{secsetup}}

We consider a large $N$ CFT with a holographic dual, defined on $S^{d-1} \times [\rm time]$. We take the CFT in the ground state $|0\rangle$. The dual spacetime is AdS$_{d+1}$ in global coordinates
\be
ds^2 = -\left(1+r^2\right) dt^2 
+{ dr^2 \over 1+r^2} + r^2 d\Omega_{d-1}^2 
\ee
We are working in units where both the radius of $S^{d-1}$ and of AdS$_{d+1}$ are set to 1.

\begin{figure}[!h]
\begin{center}
  \includegraphics[width=0.5\textwidth]{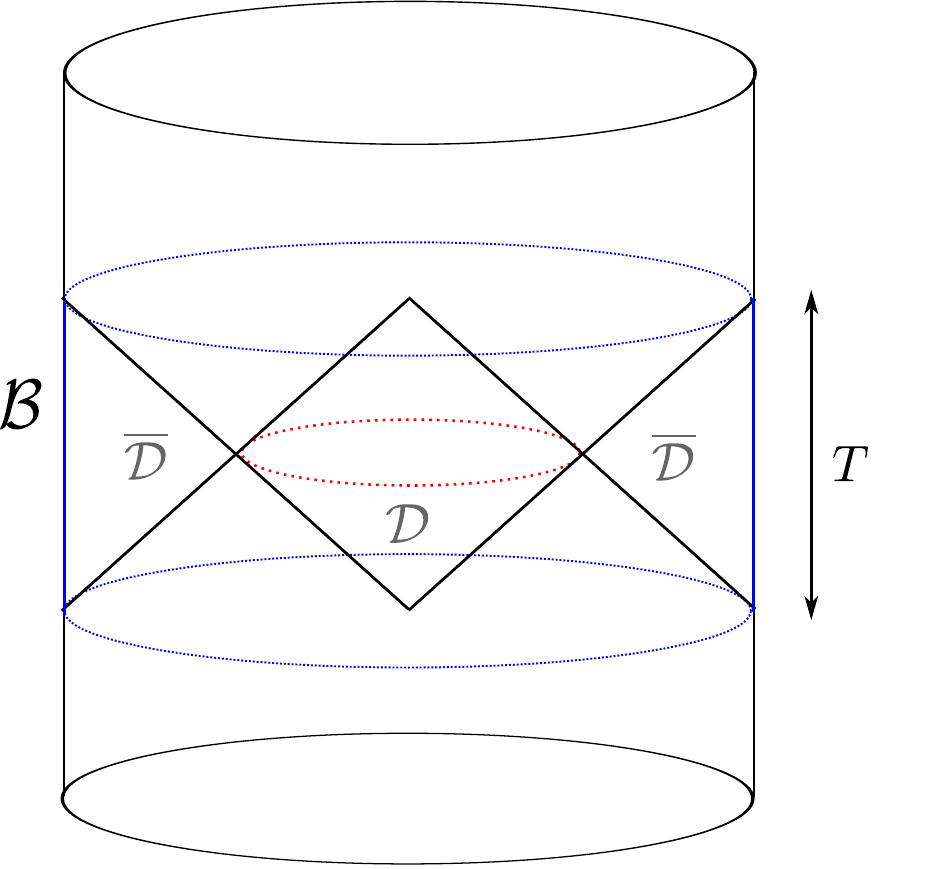}
  \caption{\em The $\tband$ ${\cal B}$ of length $T<\pi$ on the boundary of AdS spacetime, the diamond shaped region $\D$ in the bulk and its causal complement,  the annular region $\tD$.}
  \label{diamond}
\end{center}
\end{figure}

We consider a $\tband$ $\tbb$ in the CFT that is defined to be  the set of points $S^{d-1} \times \left[0, T\right]$. We are interested in the case
where this $\band$ is short, and in particular shorter than the light crossing time in AdS: $T < \pi $.

The bulk points that are causally disconnected from boundary points in $\tbb$ constitute a causal diamond $\D$ in the center of AdS as depicted in figure \ref{diamond}. The base of the diamond extends in the $r$ coordinate up to
\be
r_d = \tan\left[{\pi-T \over 2}\right]
\ee
As we make the length $T$ of the $\band$ longer, the diamond gets smaller in size and for $T \geq \pi$ the diamond disappears altogether. The causal complement of the diamond
defines an  annular domain in the bulk that is denoted as $\tD$ in figure \ref{diamond}.

If we were dealing with a non-gravitational QFT on a fixed AdS background then the two domains $D,\tD$
would correspond to a decomposition of the bulk Hilbert space in two factors. Here, we are neglecting UV divergences, and so the situation would be similar to the decomposition of the Minkowski Hilbert space into two Rindler wedges. 
More precisely the operator
algebras ${\cal A}(D)$ and ${\cal A}(\tD)$ would be well defined and independent. In a theory with gravity we do not expect to have
sharply defined local observables, so these algebras will make sense only in an approximate sense, in the large $N$ limit. Our goal in this paper is precisely to explore this loss of locality.

In some sense the set-up  resembles a black hole. For the observer confined to the time band, there is a spherical
horizon located at $r=r_d$. Approximately local degrees of freedom at smaller values of $r$ are not easily accessible to this observer. However, as we show below, just as in the case of black hole complementarity, this observer can access the interior of $\D$, using sufficiently complicated operators.

While we will elaborate on this answer shortly, we first point out how local operators in ${\cal A}(\tD)$ can be related to simple operators inside the $\tband$ ${\cal B}$ on the boundary.

Consider a free scalar field $\phi$ in  $\tD$, satisfying $\left( \Box - m^2 \right) \phi = 0$. We impose normalizable boundary conditions for this field in the time band, and additionally impose the boundary condition that a suitably rescaled field tends to the boundary operator within the time band. This corresponds to
\be
\label{bandboundary}
\begin{split}
&\lim_{r \rightarrow \infty} r^{\Delta} \phi(t, r, \Omega) = \op(t, \Omega),
\end{split}
\ee
and fixes the form of $\phi$ within $\tD$. 

More precisely, we can write the bulk field within $\tD$ as
\be
\label{bandtransfer}
\phi(t, r, \Omega) = \sum_{k, \ell} \op^T_{k, \ell} e^{-i 2 \pi k t/T} \zeta_{k, \ell}(r) Y_{\ell}(\Omega) + \text{h.c},
\ee
where $Y_{\ell}(\Omega)$ are spherical harmonics, and 
\be
\op^T_{k, \ell} = {1 \over T} \int_0^T d t \int d^{d-1}\Omega \, \op(\tau, \Omega) e^{i 2 \pi k t/T} Y_{\ell}^*(\Omega).
\ee
These modes $\op^T_{k ,\ell}$ should not be confused with the global AdS modes $\op_{n, \ell}$ that appear later and are defined in \eqref{fourierdef}. Another technicality, which is irrelevant here, is that since we have defined these modes by  convoluting the boundary operators with a function that drops sharply to $0$ at the end-points $0$ and $T$,  the action of $\op^T_{k, \ell}$ creates states that have high energy tails and are non-normalizable. 

The radial modes $\zeta_{k, \ell}$ appropriate for this setting are calculated by imposing \eqref{bandboundary} and are found to be
\be
\zeta_{k, \ell}(r) = \left({r^2 \over 1 + r^2} \right)^{\omega_k \over 2} r^{-\Delta} \, _2F_1\left(\frac{1}{2} (2-d-\ell-\omega_k+\Delta),\frac{1}{2} (\ell-\omega_k+\Delta );1-\frac{d}{2}+\Delta;-\frac{1}{r^2}\right),
\ee
where $\omega_k \equiv {2 \pi k \over T}$.
Note that at large $\ell$  the hypergeometric function
in the expression for the bulk mode grows exponentially, so that
\be
\zeta_{k,\ell}(r) \underset{\ell \rightarrow \infty}{\longrightarrow} c \,\ell^{{d-1 \over 2} -\Delta} \exp\left[\ell \,{\rm arccosh}\left( { r^2+2\over 2r^2}\right)\right],
\ee
where $c$ is an $\ell-$independent constant. 

 While \eqref{bandtransfer} gives an explicit formula for the bulk to boundary map in momentum space, it  is not possible to Fourier transform this expression back to position space as a result of the exponential growth in the mode function at large $\ell$. This means we cannot write the bulk field in terms of a convolution of the boundary operator with an ordinary function in position space on the boundary. 
\[
\nexists K_{T}, \text{such~that}~\phi(t,r, \Omega) = \int_0^T dt' \int d^{d-1} \Omega' \, \op(t', \Omega') K_{T}(t', \Omega'; t, r, \Omega).
\]
 This can also be related to the existence of bulk null geodesics that do not intersect the $\tband$ ${\cal B}$ \cite{Bousso:2012mh}.  The same technical complication arises in
the reconstruction of the AdS-Rindler wedge and of the exterior of an AdS black hole. As was first explained in \cite{Papadodimas:2012aq} and then elaborated in \cite{Morrison:2014jha}, we must understand $K_{T}$ as a distribution that is integrated only against correlators in the CFT. 
Leaving aside this subtlety, the bottom line is that we expect local operators in $\tD$ to be related to simple CFT operators in the \tband.

Note that no such direct construction is possible for operators inside the diamond $\D$ within effective field theory. Except for Gauss law tails, which  appear because the energy and other conserved charges can all be measured at infinity, local operators inside $\D$ commute with simple single-trace operators within the $\tband$ on the boundary. 

One may wonder why it would not be possible to simply apply \eqref{bandtransfer} for points inside $\D$. While the hypergeometric function in \eqref{bandtransfer} is well defined everywhere except for $r = 0$, the field that we would obtain by means of this formal extension would not obey \eqref{bandboundary} for $t > T$. So, it would differ from the correct bulk field operator inside $\D$. The extension of the field from $\tD$ to $\D$ is not uniquely determined by boundary conditions on ${\cal B}$, since we can have solutions with support in $\D$ that are zero everywhere on ${\cal B}$; so if we restrict ourselves to single-trace operators, we need information from the boundary region $[T, \pi]$ to construct the field in $\D$. 

In section \ref{secinterioroperators}, we will write down an explicit formula for operators inside $\D$, but this requires us to go beyond the single-trace sector considered above.

\section{Operator algebras in time bands\label{secopalgebra}}

In this section, we will analyze the algebra of operators ${\cal A}(\tbb)$ more closely. We show that when this algebra is appropriately defined, then we can prove a version of the Reeh-Schlieder theorem. In this context, the theorem states that all effective field theory excitations in AdS, including those in $\D$ and those in $\tD$ can be obtained by acting with operators within $\tbb$. 

In formula \eqref{bandtransfer} we have already related individual field operators to smeared single-trace operators. Now, in ordinary QFT, we can multiply field operators to obtain other field operators. This gives rise to an algebra of local operators. We can do the same by multiplying single-trace operators on the boundary smeared with functions that have support only inside the time band. 

Before we proceed, we pause to emphasize an important point. In a general QFT, there is no sense in which we can associate an algebra of operators with a time band. The set of all operators at a given time constitutes all operators in the theory. This is evident in any Hamiltonian formulation of the theory, and is formalized by the {\em time-slice axiom} \cite{Haag:1992hx}.  

The reason that we can define an algebra corresponding to a time band for CFTs with a holographic dual is because they have a special class of operators called generalized free fields \cite{Heemskerk:2009pn, Fitzpatrick:2010zm, ElShowk:2011ag}. These are local operators with a small operator dimension with the property that their correlators factorize. We will denote such a field by $\op(t, \Omega)$. We would like to consider all kinds of fields, including tensor fields, but for simplicity we omit the tensor indices. We also assume that $\op(t,\Omega)$ is Hermitian. The condition of factorization means that 
\be
\label{factorization}
\begin{split}
\langle 0| {\cal O}(t_1, \Omega_1) \ldots {\cal O}(t_{2n}, \Omega_{2 n}) | 0 \rangle = {1 \over 2^n}&\sum_{\pi} \langle 0| {\cal O}(t_{\pi_1}, \Omega_{\pi_1}){\cal O}(t_{\pi_2}, \Omega_{\pi_2}) |0 \rangle \ldots \\ &\times \langle 0 |  {\cal O}(t_{\pi_{2n-1}}, \Omega_{\pi_{2n-1}}) {\cal O}(t_{\pi_{2n}}, \Omega_{t_{\pi_{2n}}}) |0 \rangle  + \Or[{1 \over N}],
\end{split}
\ee
where the sum is over all permutations of $(1 \ldots 2 n)$. These operators are called generalized free fields because while they share the property of factorization with perturbative free fields, they do not obey a perturbative equation of motion.

In a large $N$ gauge theory, generalized free fields are low dimension single trace conformal primary operators. For example, in N=4 Super Yang Mills theory, the operators $\tr(F^2)$ and the stress tensor, $T^{\mu \nu}$ are both generalized free fields. 

We now define {\em simple operators} to be those that can be written as low-order polynomials in generalized free fields. 
{\em Complicated operators} are those that can only be expressed as polynomials of a very high order --- this includes large multi-trace operators of dimension $\Or[N]$ and also operators that change the Hamiltonian by a small amount, or not at all, such as $P_0$. 

It is convenient to define the modes of generalized free fields on the sphere through
\begin{equation} 
\label{modegff}
\op_{\ell}(t) = \int_{S^{d-1}} d^{d-1}\Omega \,\op(t,\Omega) Y_{\ell}^*(\Omega),
\end{equation}
where $Y_{\ell}(\Omega)$ are the spherical harmonics on $S^{d-1}$. 

We now consider the set of all polynomials in these modes 
\be
\label{smallalg}
\asmall({\cal B}) =  \text{span of}\{\op_{\ell_1}(t_1), \op_{\ell_2}(t_2)\op_{\ell_3}(t_3),  \ldots ,\op_{\ell_4}(t_4)\op_{\ell_5}(t_5) \cdots \op_{\ell_{\dmax}} (t_{\dmax}) \}.
\ee
We write an element of this ``small algebra'' as $\al_{ a} \in \asmall$, where the time argument of the operators on the RHS must be localized in ${\cal B}$. 

Several comments are in order. First, we introduce a cut-off in the number of operators in the product by demanding that the degree of each polynomial should not be larger than $\dmax$. We must ensure that 
\be
\dmax \ll N.
\ee
 Hence the set
 $\asmall({\cal B})$  is an algebra in a restricted sense, since arbitrary multiplications can take us outside the set. We also limit the highest allowed angular momentum mode that can enter the algebra, $\ell \leq \ell_{\text{max}} \ll N$.

These cutoffs are important for realizing the idea of complementarity: while for simple
 operators, which are dual to effective field theory experiments in the bulk, the cutoff in the definition of $\asmall({\cal B})$ is not important,
  at a fundamental level the set $\asmall({\cal B})$ is not a closed sub-algebra of the CFT. We will continue calling this set an ``algebra'', but the reader should keep this important limitation in mind.

We now establish some key properties of this small algebra with respect to the vacuum. First, consider the Hilbert space $\heft$ of all states in EFT that
can be thought of as AdS with a small number of excitations. This is produced by acting with polynomials of generalized free fields, both inside and outside the band. 
\be
\heft =\text{span of}\{\op_{\ell_1}(t_1) |0 \rangle, \op_{\ell_2}(t_2)\op_{\ell_3} (t_3) |0 \rangle,  \ldots, \op_{\ell_4}(t_4)\op_{\ell_5}(t_5) \cdots \op_{\ell_{\dmax}} (t_{\dmax})|0\rangle \}.
\ee
The difference between \eqref{smallalg} and the expressions above is that the time coordinates now are not limited to the time band and we have $0 \leq t_i \leq \pi$. 

We will now prove an analogue of the {\em Reeh-Schlieder} theorem \cite{Haag:1992hx}  for local algebras for the small algebra defined above.
\be
\label{cyclicvector}
\heft \doteq \asmall | 0 \rangle,
\ee
i.e. the set of states obtained by acting with the small algebra $\asmall$ is dense in the full Hilbert space of effective field theory. 
This is the statement that the  the CFT  ground state $|0\rangle$ is a {\em cyclic} vector for this Hilbert space with respect to the small algebra.

 In the bulk, the property \eqref{cyclicvector} can be thought of as the version of Reeh-Schlieder theorem for region $\tD$. On the boundary this is a version of Reeh-Schlieder for finite time-domains. As
explained above, this statement is non-trivial only in a situation where we can naturally define the notion of a small algebra in a time domain, for instance in large $N$ CFTs.

We first establish this result in the large $N$ limit,  $N \rightarrow \infty$, namely the free field limit when there is no interaction. These results can easily be generalized away from the free-field limit, within perturbation theory in ${1 \over N}$ as we indicate subsequently.

Let us consider smearing $\op_{\ell}(t)$ in time by a smearing function $f$ whose support is confined inside the \tband. This defines operators $X_f$ as
\begin{equation}
\label{defxf}
X_f = \int dt f(t) \op_{\ell}(t)\qquad,\quad f(t) =0 \,\,\,\,\text{for}\,\,\,\, t \notin [0, T].
\end{equation}
These operators
can be thought of as generators of the small algebra $\asmall (\tbb)$ of simple operators in the time $\band$. We do not display the dependence of $X$ on the angular momentum quantum number $\ell$ because it will not play much of a role below. 

In the large $N$ limit, the space of simple bulk states has the structure of a Fock space. As a consequence of this, the statement \eqref{cyclicvector} can be established by simply showing that 
any single particle state can be well approximated by a state of the form $X_f|0 \rangle$. More precisely, any single particle state can be written as
\begin{equation}
\label{spstates}
 |\Psi\rangle = \int_0^{\pi} dt\, g(t)\, \op_{\ell}(t) \,|0\rangle,
\end{equation}
for an appropriate choice of the function $g(t)$. Then the claim is that 
we can find a sequence of functions $f_1, \ldots f_n, \ldots$, such that
\be
\label{largencyclic}
\lim_{n \rightarrow \infty} X_{f_n} |0 \rangle \equiv \lim_{n \rightarrow \infty} \int_0^T \op_{\ell}(t) f_n(t) |0 \rangle = \int_0^{\pi} g(t) \op_{\ell}(t) |0 \rangle.
\ee

To prove this, 
we consider its converse. If \eqref{largencyclic} did not hold, then
there would exist a non-vanishing state of the form \eqref{spstates}  orthogonal   to all states produced by operators of the form \eqref{defxf}. This in particular would require that
$\langle 0| \op_{\ell}(t) |\Psi\rangle =0$ for all $t\in [0,T]$ and for all $\ell$.  We consider the function
\begin{equation}
 \label{rsproof}
 R(t) \equiv \langle 0 |\op_{\ell}(t) |\Psi\rangle.
\end{equation}
Given the positivity of the energies in the CFT this function can be analytically continued in the ${\text{Im}[t]<0}$ half-plane. We then have a meromorphic function $R(t)$ in the lower half plane with the
property that $\lim_{{\text Im}[t] \rightarrow 0^-} R(t)= 0$ for all $\text{Re}[t]\in [0,T]$. Then the ``edge of the wedge'' theorem implies that $R(t)$ vanishes everywhere. This is
inconsistent with the assumption that $|\Psi\rangle$ was a non-vanishing state of the form \eqref{spstates}, we have thus reached a contradiction. 

Having thus established that general bulk single-particle states of the form \eqref{spstates} can be arbitrarily well approximated by states produced by operators of the form \eqref{defxf}, it is easy to show  that in the large
$N$ limit the same can be done for multi-particle bulk states by induction. Say that we have proved that all $n$-particle states can be obtained by acting with operators inside $\tbb$. Then the space of $(n+1)$-particle states is spanned by
states of the form
\be
\label{psinp1const}
|\Psi_{n+1} \rangle = \int_0^{\pi} \op_{\ell}(t) g(t) |\Psi_n \rangle,
\ee
where $|\Psi_n\rangle$ is a $n$ particle state and $g(t)$ is arbitrary as above. Once again, it is possible to construct this using 
\be
|\Psi_{n+1} \rangle = \lim_{n \rightarrow \infty} X_{f_n} |\Psi_n \rangle,
\ee
for an appropriately chosen sequence of functions $f_n$ with support in $[0, T]$. If this has not been possible we would have $\langle \Psi_n | \op_{\ell}(t) |\Psi_{n+1} \rangle = 0$ for all $t \in [0, T]$. By the edge of the wedge theorem, this would require $\langle \Psi_n| \op_{\ell}(t) |\Psi_{n+1} \rangle = 0, \forall t \in [0, \pi]$, but this is in contradiction with \eqref{psinp1const}.

This property of cyclicity is a somewhat surprising property, and we can gain intuition for it by examining the structure of the norm on the space of single-particle states. Indeed, it is clear that a general function in $[0, \pi]$ cannot be well-approximated by a function in $[0, T]$ in the usual $L^2$ norm. What allows us to approximate a state produced by a function on $[0, \pi]$ with another state produced by functions on $[0, T]$ is the structure of the norm. 

Notice  that given two such states
\be
|\Psi_1 \rangle = \int_0^{\pi} dt_1  g_1(t_1) \op_{\ell}(t_1)|0 \rangle, \quad  |\Psi_2 \rangle = \int_0^{\pi} dt_2  g_2(t_2) \op_{\ell}(t_2)  |0 \rangle,
\ee
we have
\be
\langle \Psi_1 | \Psi_2 \rangle = \int_0^{\pi} d t_1 \int_0^{\pi} d t_2  g_1^*(t_1) G_{\ell}(t_1 - t_2) g_2(t_2).
\ee
Therefore the inner-product on the Hilbert space induces a bilocal product on function space that depends on the Wightman function in the ground state
\begin{align}
 \langle 0 | \op_{\ell}(t_1) \op_{\ell'} (t_2) |0\rangle  \equiv G_{\ell}(t_1 - t_2) \delta_{\ell \ell'},
\end{align}
where the delta function in the angular momentum comes from the rotational invariance of the vacuum. 
Note that in the correlator above, we have picked an ordering in the Lorentzian theory. This is a Wightman function, and not a time-ordered correlator. 

For any given mode, it is straightforward to compute this Wightman function explicitly. For example for the modes of a scalar field of dimension $\Delta$, we have
\begin{align}
 \label{2point}
\langle 0 | \op_{\ell}(t_1) \op_{\ell'}(t_2) |0\rangle = & \frac{ 2^{2-\Delta-l}  \pi ^d\Gamma (\Delta+l)}{\Gamma \left(d/2\right) 
\Gamma \left(d/2+l\right) \Gamma (\Delta) }   \cos ^{-\Delta - \ell}[t_{12}-i \epsilon]  \\ & \times
_2F_1\left(\frac{\Delta+l}{2},\frac{\Delta+l+1}{2};\frac{d}{2}+l;\cos^{-2}[t_{12}-i \epsilon]\right)\delta_{\ell \ell'},
\end{align}
where $t_{12} = t_1-t_2$. Here we have suppressed the hopefully obvious dependence on the remaining angular momentum quantum numbers characterizing operators within
the given representations $\ell,\ell'$.

The Fourier transform, $G_{\ell}(\omega) = {1\over 2\pi}\int dt e^{i \omega t} G_{\ell}(t)$, can be obtained by expanding the 2-point function in a complete set of energy eigenstates we find
\begin{equation}
%\begin{aligned}
G_{\ell}(\omega) 
= \sum_E \delta(E - \omega) |\bra{0} \op_{\ell}(0) \ket{E}|^2.
%\end{aligned}
\end{equation}
From this we learn that on general grounds $G_{\ell}(\omega) \geq 0$. Moreover at large $N$ the function $G_{\ell}(\omega)$ has support only on single-particle states in the bulk
whose energies are $E  = \Delta + 2n +\ell$, where $\ell$ is the angular momentum of the mode. 
\begin{equation}
 G_{\ell}(\omega)  = \sum_{n=0}^\infty G_{n,\ell} \delta(\omega - \Delta - 2n - \ell).
\end{equation}
where  the coefficients $G_{n,\ell}$ are the Fourier transform of \eqref{2point} and have the form
\begin{equation}
G_{n,\ell} =  \frac{4 \pi^d \Gamma\left(\Delta + n +\ell\right)\Gamma\left(\Delta + n +1 -d/2\right)}
 {\Gamma\left(d/2 \right)\Gamma\left(n+1\right)\Gamma\left(\Delta \right) \Gamma\left(\Delta + 1 - d/2 \right) \Gamma\left(d/2+ n +\ell\right)}.
\end{equation}

The important property to notice above is that $G_{n, \ell} = 0, \forall n < 0.$ So, to reconstruct a state of the form \eqref{spstates}, using \eqref{largencyclic}, we need the sequence of functions $f_n$ to match only the positive Fourier components of $g$ and not to match the function in general. This is what allows us to create any state by acting within the time band and allows \eqref{cyclicvector} to hold. 

Although we have phrased our entire discussion within free-field theory, most of the discussion above remains unchanged in its essentials when ${1 \over N}$ corrections are added.  This is because within perturbation theory, we can write the Heisenberg operators in the ${1 \over N}$ expansion as linear combinations of the free-field operators. Note that this statement is only correct perturbatively, and in terms of the action of these operators on states with energies much less than $N$. Therefore, it is clear  that, in the perturbative approximation, the span of these operators is the same as the span of the original operators. So, we conclude vacuum remains a cyclic vector with respect to $\asmall$. 

We will write the fact that we generate the state $|\Psi \rangle$ by acting with an operator with compact support as 
\be
X_f |0 \rangle \doteq |\Psi \rangle,
\ee
where it is understood that this corresponds to taking a sequence of functions in $[0, T]$ and then taking the limit. 

The reason we are careful to write $\doteq$ instead of $=$ is that, in fact, one can show that while by acting within $[0, T]$ it is possible to approximate the state $|\Psi \rangle$ arbitrarily well, one cannot always go to the limit and obtain a state that is equal to $|\Psi \rangle$. A closely related result is that we cannot exactly annihilate the vacuum by acting with operators smeared with functions in $[0, T]$. 

More precisely, we have the result
\begin{align}
\label{separatingvector}
&\nexists f(t), \,\, f\neq 0,\quad  \text{such~that} \int_0^T f(t) \op_{\ell}(t) |0 \rangle = 0.
\end{align}
The theorem \eqref{separatingvector} is sometimes framed by stating that the vacuum is a {\em separating} vector for the small algebra. We provide two pieces of caution the reader while interpreting \eqref{separatingvector}.  First, while we cannot exactly annihilate a state from within the time band, we can get arbitrarily close to 0; Second, we note that \eqref{separatingvector} only holds for generalized free fields that do not correspond to short representations  of the conformal algebra. If we are considering generalized free fields like the stress tensor or conserved currents, then it is possible to annihilate the vacuum, as we show below.

We now prove \eqref{separatingvector}. Say that a function $f(t)$ with compact support in $[0, T]$ existed so that we could use it to annihilate the vacuum. Then we must have
\be
\label{contradictseparating}
\begin{split}
X_f |0 \rangle &= \int_0^T f(t) \op_{\ell}(t) |0 \rangle = \int_0^{\pi} f(t) 
\op_{\ell}(t) |0 \rangle = \pi \sum_{n=-\infty}^{\infty} f_{-n} \op_{n, \ell} |0 \rangle = 0,
\end{split}
\ee
where we have first used the fact that $f$ has compact support to expand the region of integration, and then used a Fourier transform, with 
\be
\label{fourierdef}
\op_{n, \ell} ={1\over \pi}  \int_0^{\pi} \op_{\ell}(t) e^{i (\Delta + \ell+ 2 n ) t}; \quad f_{n} = {1\over \pi}\int_0^{\pi} f(t) e^{i (\Delta + \ell+ 2 n ) t}.
\ee
Now, unless $\op$ belongs to a short representations of the conformal algebra, then by the state operator map the action 
of $\op_{0,0}$ on the vacuum creates the primary state of this representation while $\op_{n, \ell}|0 \rangle$ with non-positive $n$ corresponds to descendants. Since these descendants are orthogonal, we see that for \eqref{contradictseparating} to hold, we must have $f_{n} = 0, \forall n \geq 0$. But then $f$ is an analytic function in the lower $t$ plane, and since $f(t) = 0$ for $t \in [T, \pi]$, by the edge of the wedge theorem, we see that $f(t) = 0, \forall t$. This proves \eqref{separatingvector}.

This proof evidently fails for operators corresponding to short representations. Here, we can annihilate the vacuum by integrating the operator with a specific spherical harmonic and then smearing it appropriately in time so as to extract
the null descendant. For example we clearly have 
\be
\int d^{d-1} \Omega \int d t \,  T^{00}(t, \Omega) f'(t)| 0 \rangle = 0,
\ee
 where $T$ is the stress-tensor, and $f$ is a function that vanishes smoothly at the end-points $[0, T]$. Similar relations hold for other conserved currents. 

As mentioned above, a related statement is that there exist states $|\Psi\rangle \in\heft$ that can be arbitrarily well approximated but not necessarily quite attained using the operators $X_f$. More specifically, there are states  $|\Psi\rangle \in\heft$  with the property that while we can find a sequence of functions $f_1, \ldots f_n, \ldots$, such that
$
\lim_{n \rightarrow \infty} \int_0^T \op_{\ell}(t) f_n(t) |0 \rangle = |\Psi\rangle
$
we also have
\begin{equation}
 \label{noexact}
\nexists f(t), \text{such~that} \int_0^T f(t) \op_{\ell}(t) |0 \rangle = |\Psi \rangle.
\end{equation}
For example, consider a state $|\Psi\rangle$ that is a superposition of global AdS modes with a maximum $n_{\rm max}$. For such a $|\Psi\rangle$,  we can prove \eqref{noexact} just as we proved \eqref{separatingvector}. If such an $f$ existed, then $e^{i t ( \Delta + \ell +2 n_{\rm max})} f(t)$ would lead to a function that was analytic in the lower $t$-plane. But any such function that vanishes in $[T, \pi]$, must vanish everywhere, and so $f$ would be have to be $0$. This is absurd, and so $f$ cannot exist.
 
The relations  \eqref{separatingvector} and \eqref{noexact} may be understood as one important difference between the algebra of operators in $[0, T]$ and in $[0, \pi]$ at infinite $N$. In the former case, we cannot produce exact energy eigenstates or annihilate the vacuum exactly, whereas in the latter we can.

Note that both \eqref{separatingvector} and \eqref{noexact} continue to be true at finite $N$. This is because we may write the action of an operator on the vacuum at finite $N$ as
\be
\al_{a} |0 \rangle = \al_{a}^0 |0 \rangle + {1 \over N} \al_a^1 |1 \rangle + \ldots,
\ee
where $\al_{a}^0 |0 \rangle$ is the state that we would have obtained at infinite N, and the remaining terms are perturbative corrections.  Now, we see that since all the terms multiplying the powers of ${1 \over N}$ are manifestly independent of $N$, $\al_{a} |0 \rangle = 0 \Rightarrow \al_{a}^0 |0 \rangle = 0$. Since we have proved that the latter cannot happen, we conclude that perturbatively the vacuum remains a separating vector. Similarly, it is not possible to generate an exact energy eigenstate through the action of the small algebra. 

However, at finite finite $N$, the relations \eqref{separatingvector} and \eqref{noexact} are less of a distinguishing factor between $\asmall$  and the algebra of single trace operators in the interval $[0, \pi]$ since at finite $N$ we cannot annihilate the vacuum, or produce exact energy eigenstates even by considering simple operators from the larger time-range.

\subsection{Explicit Construction of Arbitrary States Using Operators in the Band \label{secexplicit}}

We now turn to the explicit construction of a sequence of functions $f_n$ that, through the operators in \eqref{defxf}, can approximate any state $|\Psi \rangle$ in the larger time band. 
Let $b_k(t)$ be a complete basis of functions with compact support in $[0, T]$, where $k = 0, 1, \ldots \infty$. Then we consider the set of trial states
\be
\label{trial}
X_{f_n} |0 \rangle = \sum_{k=0}^n \alpha_k \int_0^T \op_{\ell}(t) b_k(t) dt |0 \rangle,
\ee
and choose $\alpha_k$ to minimize 
\be
r_n = ||X_{f_n} |0 \rangle - |\Psi \rangle||^2
\ee
for each $n$. Note that as $n$ increases the value of this minimum must decreases monotonically, since at each $n$, we have the choice of obtaining the previous minimum by just taking $\alpha_n = 0$. By the theorem of cyclicity above, we also 
see also that as we take $n \rightarrow \infty$ this norm must tend to 0.

If we denote the inner-product matrix between the elements of the $b_k$ basis $G_{q k}^T$
\be
G_{q k}^T \equiv \int_0^T \, d t_1 \int_0^T d t_2 b_q^*(t_1) G(t_1 - t_2) b_k(t_2),
\ee
then to minimize $r_n$ we require
\be
\sum_k \alpha_k G^T_{q k} - \int_0^{\pi} d t_2 \int_0^T d t_1 b_q^*(t_1) g(t_2) G(t_1 - t_2) = 0, \forall q.
\ee

We note that by showing that the state is separating, we have also shown that $G_{q k}^T$ is invertible. We now denote the inverse of $G_{q k}^T$ by $I_{p q}^T$, which has the property that
\be
\sum_{q=0}^{n} I_{p q}^T G_{q k}^T = \delta_{p k},
\ee
where we remind the reader that $n$ is the length of the sequence that appears in the trial wave function \eqref{trial}.  
This allows us to solve the equation above through
\be
\label{explicitconst}
f_n(t) = \sum_{q,k=0}^n I_{q k} b_q(t) \int_0^{\pi} d t_2 \int_0^T d t_1  \, g(t_2) G(t_1 - t_2)  b_k^*(t_1).
\ee
The specific choice of the basis $b_k(t)$ may be made according to convenience and does not affect the validity of the formula above. 

The same procedure can easily be used to create multi-particle states with operators that have support only within the time band.

\section{Interior operators and precursors \label{secinterioroperators}}
We now move on to the question of how to represent CFT bulk operators, which are inside the diamond $\D$, using operators in the time band. 
From bulk effective field theory, this might seem impossible. Bulk locality implies that CFT operators in the $\tband$ should commute, up to Gauss law tails,  with operators in the diamond, 
since they are spacelike separated.  On the other hand the {\em time slice axiom} of quantum field theory implies that all CFT operators are contained in the set of operators at a given time. Therefore, if the CFT has operators that represent the interior of the diamond, they must be present in the $\tband$. (This point was also discussed in \cite{Almheiri:2014lwa}.)

Although this seems to be a contradiction, it is resolved by the fact that bulk locality is an emergent concept. Operators inside $\D$ are made up of complicated operators from the $\tband$. These complicated operators have the property that they commute with the simple operators that make up operators inside $\tD$. 

In this section, we will show how to explicitly reconstruct these complicated operators.   Although this construction clearly conflicts with bulk locality, we will show that it can be done in a straightforward manner by adding   the complicated polynomials \eqref{complicatedpoly}, that approximate  $P_0 = |0 \rangle \langle 0|$, to the algebra.

The operators inside the diamond are also called precursors \cite{Polchinski:1999yd}. This terminology arises as follows. Consider the excited state, 
\be
|\text{exc} \rangle = \exp\left[i \int_{S^{d-1}} d \Omega \phi(r=0,t={T \over 2}, \Omega) \right]| 0 \rangle,
\ee
which corresponds to the vacuum excited with an S-wave at the center of the diamond. Then, this state has the property that 
\be
\label{excitedstate}
\langle \text{exc} | \al_{\alpha} | \text{exc} \rangle = \langle 0 | \al_{\alpha} |0 \rangle, \quad  \forall \al_{\alpha} \in \asmall.
\ee
This makes it appear that an observer restricted to measuring simple operators inside the time band cannot detect the presence of this excitation. 

On the other hand, an observer who has access to the entire boundary, can simply wait till the time $t = \pi$, and detect the presence of the excitation in $|\text{exc} \rangle$ in \eqref{excitedstate}.  Below, we will directly construct $\phi(r=0, t = {T \over 2}, \Omega)$ using complicated operators from the time band. These operators are called precursors because they give us information about points deep in the bulk, before this information can causally propagate to the boundary.

Before we start, it is worth mentioning that there is a simple way to construct precursors by using operators from only the time band. This is to consider the set of operators $U(t) = e^{i H t}$, 
for all values of $t$. This fact was also emphasized by Marolf \cite{Marolf:2008mf}, who argued directly from the bulk that information present in any
Cauchy slice of the boundary could be recovered from any other slice. Note that if we have access to arbitrarily complicated operators at $t = 0$, then the Hamiltonian can be evaluated on that time-slice and we can reconstruct $U(t)$. Using $U(t)$, we
 can then reconstruct the Heisenberg operators at all values of time. However, this construction is somewhat formal, and does not give insight into the nature of the ``complicated operators'' that enter into expressions for local operators in the interior of the diamond. It also suggests that we need an infinite sequence of complicated operators, labelled by different values of $t$, to reconstruct precursors. This turns out to be unnecessary in the approach that we follow below. 

Our construction proceeds in three steps. First, we remind that reader that it is possible to write the bulk field at any point in AdS, including the center of the diamond as
\begin{equation}
\label{globalrec}
\begin{split}
\phi(t, r, \Omega) &= \sum_{n} c_{n, \ell} \op_{n, \ell} e^{-i (2 n + \ell + \Delta) t} Y_{\ell}(\Omega) \chi_{n, \ell}(r) + \text{h.c}, \\
\chi_{n, \ell}(r) &= r^\ell \left(r^2+1\right)^{-{\Delta +2n + \ell \over 2}} \, _2F_1\left(-n, - \Delta -n+{d\over 2} ;\frac{d}{2}+\ell;-r^2\right), \\
c_{n, \ell} &= \frac{\Gamma \left(\frac{1}{2} (d-2\Delta-2n )\right) \Gamma \left(\frac{1}{2} (d+2\ell+2n )\right)}{\Gamma \left(\frac{d}{2}-\Delta \right) \Gamma \left(\frac{d}{2}+\ell\right)}.
\end{split}
\end{equation}
This formula follows from the standard analysis of the bulk to boundary smearing function and we refer the reader to  \cite{Banks:1998dd,  Bena:1999jv, Hamilton:2006az, Hamilton:2005ju, Hamilton:2007wj} for details and to \cite{Papadodimas:2012aq} for a review. Notice that, unlike the expansion \eqref{bandtransfer}, the 
wave functions above do not grow as $\ell \rightarrow \infty$ and so we can write \eqref{globalrec} in position space as well.

Therefore, if we could reconstruct the operators
\be
\op_{n, \ell} = {1\over \pi}  \int_0^{\pi}  dt \int d^{d-1} \Omega   \op(t, \Omega) Y_{\ell}^*(\Omega) e^{i (n + \ell + 2 \Delta) t} 
\ee
using operators from the time band, we would be able to reconstruct the local field. 

\paragraph{\bf Projector on the Vacuum \\}
We now show that to reconstruct $\op_{n, \ell}$,  we need to add only one operator to the algebra to obtain precursors. This is the operator 
\be
P_0 = |0 \rangle \langle 0|.
\ee
This operator can clearly be constructed using local operators in the CFT within the time band $[0, T]$. For example, we could write
\be
\label{p0rep}
P_0 = \lim_{\alpha \rightarrow \infty} e^{-\alpha H},
\ee
where $H$ is the CFT Hamiltonian that is simply obtained by integrating the local stress energy tensor
\be
H = \int T^{00}(t, \Omega)  d^{d-1} \Omega  - E_0,
\ee
and shifted by a constant, $E_0$,  to ensure that the ground state is annihilated by $H$. 

But, it is important, that for our purposes we do not need the exact operator $\eqref{p0rep}$ but any approximation of the form 
\be
\label{expansionp0}
{\cal P}_{\alpha, p_c} = \sum_{p=0}^{p_c} {(-1)^p (\alpha H)^p \over p!},
\ee
will also suffice provided we take $\alpha$ and $p_c$ to be large enough. 

To see how large these values have to be, note that we must take $\alpha$ large enough so that, if $|E_{\text{min}} \rangle$ is the lowest energy state above the vacuum,  we have $|P_{\alpha, p_c} |E_{\text{min}} \rangle|^2 \ll 1$. This requires $e^{-\alpha E_{\text{min}}} \ll 1$. Now, since every holographic theory contains the graviton, the first excited state has an energy that cannot be larger than $d$, $E_{\text{min}}  \leq d$.   Second, consider the highest energy state $E_{\text{max}}$ for states in $\heft$. We must keep enough terms in the polynomial to ensure that \eqref{expansionp0} is a good approximation to the exponential for this state as well. This implies that that the two conditions on the cutoffs are
\be
\label{cutoffcondition}
e^{-\alpha d} \ll 1, \quad p_c \gg \alpha E_{\text{max}}.
\ee

We expect that $E_{\text{max}} \ll N$, since for states with higher energy than this, our description of the physics in terms of generalized free fields breaks down.  Moreover, since we are working at leading order in $N$, it is sufficient to suppress the lowest excited state by a factor of ${1 \over N}$. Therefore, one choice of cutoffs that meets the condition \eqref{cutoffcondition} is 
$\alpha = \ln(N)$ and $p_c = N \ln(N)$. The reader may choose to work with different cutoffs provided that \eqref{cutoffcondition} is satisfied. 

It is important to understand that no choice of cutoffs on $\heft$ will allow us to include a good approximation to $P_0$ within our simple algebra. For example, let us say we attempt to take $p_c = \dmax$ to include an expansion of the form \eqref{expansionp0} in the algebra, where we remind the reader that $\dmax$ is the largest allowed degree of a polynomial in the simple algebra.  But now we see that $E_{\text{max}} \geq d \dmax$, and for states with this energy, and the cutoff $p_c = \dmax$, the expansion \eqref{expansionp0} does not approximate $P_0$ well. 

\paragraph{\bf Construction of $\op_{n, \ell}$ \\}
We now show how to use $P_0$, or alternately the polynomials ${\cal P}_{\alpha, p_c}$ from \eqref{expansionp0}, to construct $\op_{n, \ell}$ 

As we have discussed, at large $N$,  the Hilbert space has the structure of a Fock space. In this limit, we introduce a natural basis of states for $\heft$ by writing
\be
\label{fockstate}
|p_{n_1,\ell_1} \ldots p_{n_j, \ell_j} \ldots \rangle = \prod_{j=0}^{\dmax} \left(\Gamma(p_{n_j, \ell_j} +1) G_{n_j, \ell_j}  \right)^{-{1 \over 2}} \left(\op_{n_j, \ell_j} \right)^{p_{n_j, \ell_j}}  | 0 \rangle,
\ee
where the product above ranges over all allowed descendants of the primary operator $\op$, which are limited by $\dmax$ by our cutoff above. 
In this basis, the operator $\op_{n, \ell}$ has the natural simple harmonic oscillator form
\be
\op_{n, \ell} = \sum_{\{p_{n_j, \ell_j}\}} \sqrt{p_{n, \ell} G_{n, \ell}} |p_{0,0} \ldots p_{n, \ell}-1 \ldots \ldots \rangle \langle p_{0,0}, \ldots p_{n, \ell} \ldots |,
\ee
where the sum ranges over all allowed $p_j$. 

But note that we already know how to construct the states in the Fock space using operators in the $\tband$. Let us introduce some notation to represent this. We denote the operator $X \in \asmall$ that creates the state \eqref{fockstate} by
\be
X[p_{n_1, \ell_1} \ldots p_{n_j, \ell_j} \ldots] |0 \rangle \doteq |p_{n_1,\ell_1} \ldots p_{n_j, \ell_j} \ldots \rangle.
\ee

With this notation, we see that the mode of the boundary operator can be written as
\be
\op_{n, \ell} \doteq  \sum_{\{p_{n_j, \ell_j}\}} \sqrt{p_{n, \ell} G_{n, \ell}} X[p_{0,0} \ldots p_{n, \ell}-1 \ldots] P_0 X[p_{0,0}  \ldots p_{n, \ell} \ldots ]^{\dagger}.
\ee
We can now simply write the field operator as
\be
\label{precursor}
\begin{split}
\phi(t, r, \Omega) \doteq \sum_{n, \ell} \sum_{\{p_{n_j, \ell_j}\}} &\sqrt{p_{n, \ell} G_{n, \ell}} X[p_{0,0} \ldots p_{n, \ell}-1 \ldots] P_0 X[p_{0,0} \ldots p_{n, \ell} \ldots ]^{\dagger}  \\ &\times c_{n, \ell} \chi_{n, \ell}(r) e^{-i(\Delta + 2 n + \ell) t} Y_{\ell}(\Omega) + \text{h.c}.
\end{split}
\ee
We remind the reader that we use $\doteq$ because the operators in the time band can reproduce a given state only in the limit shown in \eqref{largencyclic}. Except for $P_0$, all the other operators that appear above explicitly belong to the simple algebra and $P_0$ itself can be obtained as a limit of a sequence of polynomials ${\cal P}_{\alpha, p_c}$. 

Note that to obtain this result, we had to use the important result of cyclicity from the discussion above: the set of states obtained by acting with the small algebra is dense in the full Hilbert space. Therefore any state obtained by the action of $\op_{n, \ell}$ on the vacuum can also be obtained by the action of an appropriate operator from the time band.

The statement of cyclicity by itself does not allow us to represent the operator $\op_{n, \ell}$ in terms of operators from the time band. Rather, it tells us about its action on the state $|0 \rangle$. By inserting the projector $P_0 = |0 \rangle \langle 0 |$ and sandwiching it between a sequence of operators $X[p_{0,0} \ldots]$ above, we are able to reproduce the entire operators $\op_{n, \ell}$, and in turn the local field in the interior of the diamond.

We believe that the expression \eqref{precursor} gives a remarkably simple expression for precursors in terms of boundary operators.

\section{Conclusion \label{secconclusion}}
In this paper, we have essentially focused on two results. One of them is
that large-scale non-locality is an essential feature of quantum-gravity. Such a loss of causality is an ingredient in proposals of black-hole complementarity. In previous work, two of us showed  that several recent versions of the information paradox could be resolved by accepting that simple local operators at a point in the interior of the black hole could be identified with complicated operators near the boundary of AdS that were causally disconnected from that point. 

Here, we see that this phenomenon is evident in empty AdS. 
In particular, our formula \eqref{precursor} shows that one can write the field  at a point in the center of AdS purely in terms of a complicated polynomial of operators in a time band $T < \pi$, even though all of these are causally disconnected from the center of the AdS.

A related question has to do with the approximation in which locality arises within effective field theory. Here, we argued that the correct way to understand this is in terms of an approximate ``algebra'' of simple operators in the time band. It is possible to define such an algebra, only in the large $N$ limit where there is a parametric separation between light generalized free fields and more complicated multi-trace operators comprised of polynomials of $N$ such fields. We showed that, if one makes a distinction between simple and complicated operators, then simple operators in a time band of width $T < \pi$ on the boundary obey a version of the Reeh-Schlieder theorem. This is the dual of the Reeh-Schlieder theorem for the region in the bulk that is causally connected to this time band. 

One interesting question, which we hope to explore further has to do with whether it is possible to define a natural modular Hamiltonian for the time-band.  This can be done entirely algebraically on the boundary using the techniques of Tomita-Takesaki theory  as follows.  We can define the following anti-linear operator
\be
S \al_{\alpha} |0 \rangle = A_{\alpha}^{\dagger} |0 \rangle, \quad \al_{\alpha} \in \asmall,
\ee
which is well defined on $\heft$ because the vacuum is cyclic and separating with respect to $\asmall$. Then the modular Hamiltonian can be defined through $H_{\text{mod}} = -\log(S^{\dagger} S)$. It would be interesting to check that this should correspond to the bulk modular Hamiltonian for $\tD - D$.

Precursors have been discussed extensively in the literature. In \cite{Susskind:1999ey}, it was proposed that Wilson loops may act as precursors building on the intuition that the bulk duals of Wilson loops are string worldsheets that extend into the bulk, and may therefore detect excitations in the interior.  Although this is a natural guess, as pointed out \cite{Giddings:2001pt}, this is incorrect. Wilson loops are dual to operators that create a perturbative string excitation in the bulk. To the extent that perturbative string theory is local, Wilson loop operators in a region of the boundary cannot detect information from a causally disconnected region in the bulk. In the presence of such an excitation, Wilson loops cannot be computed through a minimal area surface any longer. As explained in \cite{Jafferis:2014lza}, this is not dissimilar to the fact that, in the vacuum, correlation functions can be computed in a geodesic approximation. However, while bulk geodesics do respond to bulk excitations, this does not imply that boundary correlators are precursors; all that happens is that the duality between correlators and geodesics duality breaks down in the presence of a bulk excitation.

It is now believed, based on the HRT formula for the entanglement entropy of a region \cite{Hubeny:2007xt} that the entanglement entropy does provide an example of a precursor that is sensitive to bulk dynamics in the ``entanglement wedge'' of the region.  
A related proposal was made in \cite{Jafferis:2014lza}, where it was proposed that the modular Hamiltonian for a region was dual to the Area operator for the minimal bulk surface.  As we mentioned earlier, an important difference between previous work and our study is that we are considering a boundary region that has no non-trivial causal complement. Nevertheless, locality emerges when we focus on simple operators, whereas non-local information is stored in complex polynomials. 

It is rather remarkable that we are able to construct a toy model of black hole complementarity in this simple setting. In empty AdS, we can only examine non-locality on length scales that are the AdS radius. We believe that, in their essentials, these ideas should also apply to flat-space black holes, where we require non-locality over the the distance that radiation travels in the evaporation time of the black hole. However, it would be very interesting to understand this in greater detail.

\section*{Acknowledgments}		

We would like to thank Jan de Boer, Borun Chowdhury, Rajesh Gopakumar,  R. Loganayagam, Shiraz Minwalla, Juan Maldacena, Erik Verlinde for useful discussions and Daniel Jafferis for discussions and comments on a draft of this manuscript. We are also grateful to all members of the string theory groups at CERN, Groningen, ICTS and the Indian Institute of Science for several useful discussions.

\bibliographystyle{JHEPmod}
\bibliography{references}

\providecommand{\href}[2]{#2}\begingroup\raggedright\begin{thebibliography}{10}

\bibitem{Mathur:2009hf}
S.~D. Mathur, {\it {The Information paradox: A Pedagogical introduction}},
  {\em Class.Quant.Grav.} {\bf 26} (2009) p.~224001,
  [\href{http://xxx.lanl.gov/abs/0909.1038}{{\tt arXiv:0909.1038}}].

\bibitem{Almheiri:2012rt}
A.~Almheiri, D.~Marolf, J.~Polchinski, and J.~Sully, {\it {Black Holes:
  Complementarity or Firewalls?}},  {\em JHEP} {\bf 1302} (2013) p.~062,
  [\href{http://xxx.lanl.gov/abs/1207.3123}{{\tt arXiv:1207.3123}}].

\bibitem{Almheiri:2013hfa}
A.~Almheiri, D.~Marolf, J.~Polchinski, D.~Stanford, and J.~Sully, {\it {An
  Apologia for Firewalls}},  {\em JHEP} {\bf 1309} (2013) p.~018,
  [\href{http://xxx.lanl.gov/abs/1304.6483}{{\tt arXiv:1304.6483}}].

\bibitem{Marolf:2013dba}
D.~Marolf and J.~Polchinski, {\it {Gauge/Gravity Duality and the Black Hole
  Interior}},  {\em Phys.Rev.Lett.} {\bf 111} (2013) p.~171301,
  [\href{http://xxx.lanl.gov/abs/1307.4706}{{\tt arXiv:1307.4706}}].

\bibitem{'tHooft:1984re}
G.~'t~Hooft, {\it {On the Quantum Structure of a Black Hole}},  {\em
  Nucl.Phys.} {\bf B256} (1985) p.~727.

\bibitem{Susskind:1993if}
L.~Susskind, L.~Thorlacius, and J.~Uglum, {\it {The Stretched horizon and black
  hole complementarity}},  {\em Phys.Rev.} {\bf D48} (1993) pp.~3743--3761,
  [\href{http://xxx.lanl.gov/abs/hep-th/9306069}{{\tt hep-th/9306069}}].

\bibitem{Papadodimas:2013jku}
K.~Papadodimas and S.~Raju, {\it {State-Dependent Bulk-Boundary Maps and Black
  Hole Complementarity}},  {\em Phys.Rev.} {\bf D89} (2014), no.~8 p.~086010,
  [\href{http://xxx.lanl.gov/abs/1310.6335}{{\tt arXiv:1310.6335}}].

\bibitem{Papadodimas:2013wnh}
K.~Papadodimas and S.~Raju, {\it {The Black Hole Interior in AdS/CFT and the
  Information Paradox}},  {\em Phys.Rev.Lett.} {\bf 112} (2014), no.~5
  p.~051301, [\href{http://xxx.lanl.gov/abs/1310.6334}{{\tt arXiv:1310.6334}}].

\bibitem{Papadodimas:2015jra}
K.~Papadodimas and S.~Raju, {\it {Comments on the Necessity and Implications of
  State-Dependence in the Black Hole Interior}},
  \href{http://xxx.lanl.gov/abs/1503.0882}{{\tt arXiv:1503.0882}}.

\bibitem{Papadodimas:2015xma}
K.~Papadodimas and S.~Raju, {\it {Local Operators in the Eternal Black Hole}},
  \href{http://xxx.lanl.gov/abs/1502.0669}{{\tt arXiv:1502.0669}}.

\bibitem{Maldacena:1997re}
J.~M. Maldacena, {\it {The Large N limit of superconformal field theories and
  supergravity}},  {\em Adv.Theor.Math.Phys.} {\bf 2} (1998) pp.~231--252,
  [\href{http://xxx.lanl.gov/abs/hep-th/9711200}{{\tt hep-th/9711200}}].

\bibitem{Witten:1998qj}
E.~Witten, {\it {Anti-de Sitter space and holography}},  {\em Adv. Theor. Math.
  Phys.} {\bf 2} (1998) pp.~253--291,
  [\href{http://xxx.lanl.gov/abs/hep-th/9802150}{{\tt hep-th/9802150}}].

\bibitem{Gubser:1998bc}
S.~Gubser, I.~R. Klebanov, and A.~M. Polyakov, {\it {Gauge theory correlators
  from noncritical string theory}},  {\em Phys.Lett.} {\bf B428} (1998)
  pp.~105--114, [\href{http://xxx.lanl.gov/abs/hep-th/9802109}{{\tt
  hep-th/9802109}}].

\bibitem{Balasubramanian:2013lsa}
V.~Balasubramanian, B.~D. Chowdhury, B.~Czech, J.~de~Boer, and M.~P. Heller,
  {\it {Bulk curves from boundary data in holography}},  {\em Phys. Rev.} {\bf
  D89} (2014), no.~8 p.~086004, [\href{http://xxx.lanl.gov/abs/1310.4204}{{\tt
  arXiv:1310.4204}}].

\bibitem{Myers:2014jia}
R.~C. Myers, J.~Rao, and S.~Sugishita, {\it {Holographic Holes in Higher
  Dimensions}},  {\em JHEP} {\bf 06} (2014) p.~044,
  [\href{http://xxx.lanl.gov/abs/1403.3416}{{\tt arXiv:1403.3416}}].

\bibitem{Headrick:2014eia}
M.~Headrick, R.~C. Myers, and J.~Wien, {\it {Holographic Holes and Differential
  Entropy}},  {\em JHEP} {\bf 10} (2014) p.~149,
  [\href{http://xxx.lanl.gov/abs/1408.4770}{{\tt arXiv:1408.4770}}].

\bibitem{Marolf:2008mg}
D.~Marolf, {\it {Holographic Thought Experiments}},  {\em Phys. Rev.} {\bf D79}
  (2009) p.~024029, [\href{http://xxx.lanl.gov/abs/0808.2845}{{\tt
  arXiv:0808.2845}}].

\bibitem{Marolf:2008mf}
D.~Marolf, {\it {Unitarity and Holography in Gravitational Physics}},  {\em
  Phys.Rev.} {\bf D79} (2009) p.~044010,
  [\href{http://xxx.lanl.gov/abs/0808.2842}{{\tt arXiv:0808.2842}}].

\bibitem{Mintun:2015qda}
E.~Mintun, J.~Polchinski, and V.~Rosenhaus, {\it {Bulk-Boundary Duality, Gauge
  Invariance, and Quantum Error Corrections}},  {\em Phys. Rev. Lett.} {\bf
  115} (2015), no.~15 p.~151601, [\href{http://xxx.lanl.gov/abs/1501.0657}{{\tt
  arXiv:1501.0657}}].

\bibitem{Almheiri:2014lwa}
A.~Almheiri, X.~Dong, and D.~Harlow, {\it {Bulk Locality and Quantum Error
  Correction in AdS/CFT}},  {\em JHEP} {\bf 04} (2015) p.~163,
  [\href{http://xxx.lanl.gov/abs/1411.7041}{{\tt arXiv:1411.7041}}].

\bibitem{Giddings:2015lla}
S.~B. Giddings, {\it {Hilbert space structure in quantum gravity: an algebraic
  perspective}},  {\em JHEP} {\bf 12} (2015) p.~099,
  [\href{http://xxx.lanl.gov/abs/1503.0820}{{\tt arXiv:1503.0820}}].

\bibitem{Papadodimas:2012aq}
K.~Papadodimas and S.~Raju, {\it {An Infalling Observer in AdS/CFT}},  {\em
  JHEP} {\bf 1310} (2013) p.~212,
  [\href{http://xxx.lanl.gov/abs/1211.6767}{{\tt arXiv:1211.6767}}].

\bibitem{Headrick:2014cta}
M.~Headrick, V.~E. Hubeny, A.~Lawrence, and M.~Rangamani, {\it {Causality \&
  holographic entanglement entropy}},  {\em JHEP} {\bf 12} (2014) p.~162,
  [\href{http://xxx.lanl.gov/abs/1408.6300}{{\tt arXiv:1408.6300}}].

\bibitem{Jafferis:2015del}
D.~L. Jafferis, A.~Lewkowycz, J.~Maldacena, and S.~J. Suh, {\it {Relative
  entropy equals bulk relative entropy}},
  \href{http://xxx.lanl.gov/abs/1512.0643}{{\tt arXiv:1512.0643}}.

\bibitem{Dong:2016eik}
X.~Dong, D.~Harlow, and A.~C. Wall, {\it {Bulk Reconstruction in the
  Entanglement Wedge in AdS/CFT}},
  \href{http://xxx.lanl.gov/abs/1601.0541}{{\tt arXiv:1601.0541}}.

\bibitem{Freivogel:2016zsb}
B.~Freivogel, R.~A. Jefferson, and L.~Kabir, {\it {Precursors, Gauge
  Invariance, and Quantum Error Correction in AdS/CFT}},
  \href{http://xxx.lanl.gov/abs/1602.0481}{{\tt arXiv:1602.0481}}.

\bibitem{Morrison:2014jha}
I.~A. Morrison, {\it {Boundary-to-bulk maps for AdS causal wedges and the
  Reeh-Schlieder property in holography}},  {\em JHEP} {\bf 1405} (2014)
  p.~053, [\href{http://xxx.lanl.gov/abs/1403.3426}{{\tt arXiv:1403.3426}}].

\bibitem{Bousso:2012mh}
R.~Bousso, B.~Freivogel, S.~Leichenauer, V.~Rosenhaus, and C.~Zukowski, {\it
  {Null Geodesics, Local CFT Operators and AdS/CFT for Subregions}},
  \href{http://xxx.lanl.gov/abs/1209.4641}{{\tt arXiv:1209.4641}}.

\bibitem{Haag:1992hx}
R.~Haag, {\em {Local quantum physics: Fields, particles, algebras, 2nd ed.}}
\newblock Springer, 1992.

\bibitem{Heemskerk:2009pn}
I.~Heemskerk, J.~Penedones, J.~Polchinski, and J.~Sully, {\it {Holography from
  Conformal Field Theory}},  {\em JHEP} {\bf 0910} (2009) p.~079,
  [\href{http://xxx.lanl.gov/abs/0907.0151}{{\tt arXiv:0907.0151}}].

\bibitem{Fitzpatrick:2010zm}
A.~Fitzpatrick, E.~Katz, D.~Poland, and D.~Simmons-Duffin, {\it {Effective
  Conformal Theory and the Flat-Space Limit of AdS}},  {\em JHEP} {\bf 1107}
  (2011) p.~023, [\href{http://xxx.lanl.gov/abs/1007.2412}{{\tt
  arXiv:1007.2412}}].

\bibitem{ElShowk:2011ag}
S.~El-Showk and K.~Papadodimas, {\it {Emergent Spacetime and Holographic
  CFTs}},  \href{http://xxx.lanl.gov/abs/1101.4163}{{\tt arXiv:1101.4163}}.

\bibitem{Polchinski:1999yd}
J.~Polchinski, L.~Susskind, and N.~Toumbas, {\it {Negative energy,
  superluminosity and holography}},  {\em Phys. Rev.} {\bf D60} (1999)
  p.~084006, [\href{http://xxx.lanl.gov/abs/hep-th/9903228}{{\tt
  hep-th/9903228}}].

\bibitem{Banks:1998dd}
T.~Banks, M.~R. Douglas, G.~T. Horowitz, and E.~J. Martinec, {\it {AdS dynamics
  from conformal field theory}},
  \href{http://xxx.lanl.gov/abs/hep-th/9808016}{{\tt hep-th/9808016}}.

\bibitem{Bena:1999jv}
I.~Bena, {\it {On the construction of local fields in the bulk of AdS(5) and
  other spaces}},  {\em Phys.Rev.} {\bf D62} (2000) p.~066007,
  [\href{http://xxx.lanl.gov/abs/hep-th/9905186}{{\tt hep-th/9905186}}].

\bibitem{Hamilton:2006az}
A.~Hamilton, D.~N. Kabat, G.~Lifschytz, and D.~A. Lowe, {\it {Holographic
  representation of local bulk operators}},  {\em Phys.Rev.} {\bf D74} (2006)
  p.~066009, [\href{http://xxx.lanl.gov/abs/hep-th/0606141}{{\tt
  hep-th/0606141}}].

\bibitem{Hamilton:2005ju}
A.~Hamilton, D.~N. Kabat, G.~Lifschytz, and D.~A. Lowe, {\it {Local bulk
  operators in AdS/CFT: A Boundary view of horizons and locality}},  {\em
  Phys.Rev.} {\bf D73} (2006) p.~086003,
  [\href{http://xxx.lanl.gov/abs/hep-th/0506118}{{\tt hep-th/0506118}}].

\bibitem{Hamilton:2007wj}
A.~Hamilton, D.~N. Kabat, G.~Lifschytz, and D.~A. Lowe, {\it {Local bulk
  operators in AdS/CFT and the fate of the BTZ singularity}},
  \href{http://xxx.lanl.gov/abs/0710.4334}{{\tt arXiv:0710.4334}}.

\bibitem{Susskind:1999ey}
L.~Susskind and N.~Toumbas, {\it {Wilson loops as precursors}},  {\em Phys.
  Rev.} {\bf D61} (2000) p.~044001,
  [\href{http://xxx.lanl.gov/abs/hep-th/9909013}{{\tt hep-th/9909013}}].

\bibitem{Giddings:2001pt}
S.~B. Giddings and M.~Lippert, {\it {Precursors, black holes, and a locality
  bound}},  {\em Phys.Rev.} {\bf D65} (2002) p.~024006,
  [\href{http://xxx.lanl.gov/abs/hep-th/0103231}{{\tt hep-th/0103231}}].

\bibitem{Jafferis:2014lza}
D.~L. Jafferis and S.~J. Suh, {\it {The Gravity Duals of Modular
  Hamiltonians}},  \href{http://xxx.lanl.gov/abs/1412.8465}{{\tt
  arXiv:1412.8465}}.

\bibitem{Hubeny:2007xt}
V.~E. Hubeny, M.~Rangamani, and T.~Takayanagi, {\it {A Covariant holographic
  entanglement entropy proposal}},  {\em JHEP} {\bf 0707} (2007) p.~062,
  [\href{http://xxx.lanl.gov/abs/0705.0016}{{\tt arXiv:0705.0016}}].

\end{thebibliography}\endgroup
\end{document}